# Invalid Methods of Causal Inference in Physics Education Research

M. B. Weissman

*Department of Physics, University of Illinois at Urbana-Champaign*

*1110 West Green Street, Urbana, IL 61801-3080*

**Abstract**

Finding good educational policies requires sound estimates of their potential effects. Methods for making such estimates, i.e. finding causal estimands, have made great progress in the last few decades. Nevertheless, serious errors in causal reasoning have been found previously in papers n a leading physics education journal, Physical Review Physics Education Research. Here we examine three more recent papers from that journal that present explicit methods of causal inference. The methods given include major errors, including in identifying causal mediation, choosing variables to control for, and imputing missing data.

**Introduction**

Awareness is growing that modern causal inference methods (Imbens and Rubin 2015) are needed to help educators  predict the consequences of different choices in teaching methods and in educational policy(Murnane and Willett 2011).  Especially in the last four decades, sophisticated methods of causal inference have been developed to estimate such potential outcomes from observational data.(Pearl, Glymour, and Jewell 2016; Hernán and Robins 2020) Field-specific introductions to these methods are now available for education(Murnane and Willett 2011), economics(Varian 2016; 2015), , psychology(Foster 2010; Rohrer 2018), epidemiology(Hernán and Robins 2020),  biology(Glymour, Zhang, and Spirtes 2019), public health(Glass et al. 2013), sociology(Gangl 2010), political science(Keele 2015), and other fields (Imbens and Rubin 2015), with ref.(Rohrer 2018) providing a particularly accessible primer.

One might expect that physics, the most consistently mathematical of the natural sciences, would take the lead in employing valid new methods of quantitative reasoning to address its educational challenges.  Nevertheless, in a recent paper (Weissman 2021) I showed that several papers published before 2021  in the leading journal of physics education, Physical Review Physics Education Research (PRPER), made major causal inference errors in particular  applications. My previous paper focused on instances in which particular causal patterns were assumed without regard to the wide variety of plausible causal patterns equally consistent with the data. I included a very brief introduction(Weissman 2021) to the diagrammatic methods by which causal relations are often represented, (Pearl, Glymour, and Jewell 2016; Hernán and Robins 2020; Rohrer 2018), which need not be repeated here.

Different specific issues are raised by three more recent PRPER papers, published in 2021. (Walsh et al. 2021; Young and Caballero 2021; Verostek, Miller, and Zwickl 2021) These papers include not only examples but explicit methods instructions whose use would lead to seriously biased causal estimands. The issues raised include the rules for which variables should be included or excluded from regression models to obtain unbiased causal estimands, the ways to identify causal mediators, the specification of  potential outcomes, and the ways to impute missing data. The errors seriously distort causal implications for at least one major policy decision, the use of standardized tests in admissions decisions.

The central conclusion of this paper will be to reinforce my call for improving awareness of causal inference methods throughout the PER field. My arguments here involve no new methods but are just applications of the standard beginning methods of the field. Some of the introductory material included may be too elementary for many readers, but it is intended to make the arguments accessible to PER workers. PRPER has expressed a reluctance to publish any further work that describes some of its papers as "incorrect", indicating that the field may be in need of some outside intervention, to which this paper is meant to contribute.

## **Background on mediation, bias from including inappropriate variables, and treatment of missing data**

Within the potential outcome or counterfactual approaches to causation, the average causal effect of a change in one variable on another variable is the average change in the second variable that would occur if one were to change the first variable without intervening to change anything else. (Pearl, Glymour, and Jewell 2016; Hernán and Robins 2020) This definition has an obvious correspondence to part of what one needs to know to decide whether to make the change in question, although for a more accurate estimation of the utility of some action one should know the full change in the probability distribution for the outcome rather than just its mean. A causal estimand is a formula estimating that causal effect from various correlations among other variables, including the cause being studied. (Pearl, Glymour, and Jewell 2016; Hernán and Robins 2020)

A causal estimand is said to be biased if it systematically mis-estimates the effect. Bias can arise when the estimand either contains terms that do not correspond to results of intervening on the cause or omits relevant terms that do correspond to such changes. (An excellent easily readable introduction to this issue can be found in ref. (Rohrer 2018).) The different coefficients found in different predictive models employing different predictive variables are not themselves "biased"; they are simply the coefficients of different models that would be used with different available predictive information. They become biased only if they are used to estimate something else- the causal effects. It is meaningless to say which coefficient is "true" and which is "biased" without specifying a causal question and the true causal model.

Descriptions of bias in causal estimands often focus on the role of omitted variables (called confounders) that have effects on both the cause in question and the outcome. (Rohrer 2018) In a diagram, the simplest confounder would be one with arrows to both the cause and the outcome. Simple cause-outcome correlations in models omitting the confounders give biased estimates of the actual causal effects because they include correlations between the cause and the outcome induced by the confounders. Controlling for confounders, e.g. by including them in an appropriate regression model, can remove this bias.

A causal mediator is a variable which appears as a node on a path from a cause to an outcome. Tar deposits in lungs provide a classic example, because tar appears on an important pathway from smoking to lung cancer. A more topical example would be inflammation as a mediator on one path from SARS-CoV-2 infection to death.(The_RECOVERY_Collaborative_Group 2021) Operationally, a mediator can be identified because the portion of the causal effect that occurs via the mediated path can be shut down if the mediator can somehow be prevented from changing despite changes in the earlier cause.

This operational definition of mediation is a reminder of why it can sometimes be important to identify mediators. Sometimes it is easier to change the mediators by some intervention than it is to change the preceding cause. For example, intervening on the SARS-CoV-2→ death pathway by suppressing the inflammation mediator with dexamethasone is fairly effective (The_RECOVERY_Collaborative_Group 2021), although not nearly as effective as suppressing Covid itself with vaccines would have been. The effect of this intervention provides strong evidence that inflammation is indeed an important mediator, not just a marker for actual causes, a result that could not be determined simply from previously observed correlations. For a contrasting example, the weaker effectiveness of steroids in reducing mortality from sepsis shows that the inflammation accompanying sepsis is not such an important mediator in that analogous case.(Yao et al. 2019)

When it is suspected that one of two variables mediates the other's effects on the outcome, the time order of the two potential causes can eliminate one of the two possible mediation relations, since the mediator must came after its cause. Mere time order does not, however, establish that a mediation relation exists. When, as often happens, the observed variables are not themselves on the causal path but rather are measurements of underlying latent variables, the time order of the measurements does not even constrain the direction of any possible causal relation between those latent variables. For example, if someone measures that they have a fever and then gets a PCR test that is positive for SARS-CoV-2, it would not be correct to use the time order of the variables temperature measurement and viral measurement to conclude that the fever caused the viral infection. The starting time order of the latent variables temperature and infection is opposite to the time order of their measurements, and in this instance the latent time order gives the true causal direction.

The portion of a causal effect that occurs via a measured mediator is often termed an "indirect effect". Any remaining causal effect for which no mediator has been identified is then termed the "direct effect". Despite this terminology, it is important to realize that there is no fundamental difference between these types of causation. On the classical scale, all causation proceeds through time-like event paths. Therefore any classical effect can be converted from "direct" to "indirect" by recording and controlling for a big enough set of mediators. Thus, despite the infelicitous standard terminology, the description of a causal effect as "indirect" should not be interpreted to mean that it is any less real than one described as "direct".

In general causal estimates of the magnitude of e.g. smoking→cancer or SARS-CoV-2→death will be severely biased if they are based on models in which one controls for mediators, e.g. by including them in a linear regression model.(Rohrer 2018) The effect of smoking cigarettes on lung cancer rates should not be estimated by excluding the large portion of the effect that occurs via tar deposits, since the tar deposits are themselves usually caused by the smoking. We say SARS-CoV-2 causes death, even though if one controlled for known mediators such as low blood oxygen, lung scarring, blood clots, and systemic inflammation, the "direct effect" would be small.

Mediators are not the only variables that should not be controlled for if one wishes to use model coefficients to obtain an unbiased causal estimand, although they may be the most obvious such type of variable. Estimands will also be biased by controlling for variables that are affected both by the suspected cause and by other unmeasured causes that have effects on the outcome. Such variables are called colliders because in a causal diagram arrows from the suspected cause and the unmeasured causes collide on them. Controlling for a collider, e.g. by inclusion in a regression model, thus introduces collider stratification bias to the estimand. (Pearl, Glymour, and Jewell 2016; Greenland 2003; Rohrer 2018)

One of the best-known illustrations of collider bias is the low birth weight paradox. (VanderWeele 2014) In a model controlling for birth weight, which is a risk factor for infant mortality, maternal smoking becomes *negatively* associated with infant mortality. This does not mean that smoking is protective, but rather that the model was biased. Out-of-model causes of low birth weight (e.g. certain diseases or types of drug use) cause low birth weight, as does smoking or some variable associated with smoking. Thus birth weight is a collider between smoking and those out-of-model causes. A non-smoker with a low birth weight infant is more likely to have those other causes than is a mother whose smoking already created a likelihood of low birth weight. It turns out that the other causes are more dangerous than smoking, so among low birth weight infants those with non-smoking mothers are less likely to survive. Omitting birthweight from the model gives a positive association between smoking and infant mortality, which gives a more accurate representation of the causal effect. (VanderWeele 2014)

In many real-world observational studies, some of the data are missing. It is often advantageous to impute the probability distributions of the missing data using the other data, both to avoid throwing out information and to avoid biased estimates of the true correlations that can arise when the data are not missing completely at random. An easily readable introduction to such imputation has been published in PRPER.(Nissen, Donatello, and Van Dusen 2019)

Incorrect imputation methods can introduce bias even when the data are missing completely at random. To explain the key reason, we can consider a toy example, so simple that one would not be tempted to use imputation in practice because it adds nothing to the information available from the complete-case data set. For simplicity I will describe the logic assuming large-N samples of bivariate normal variables, X and Y with correlation coefficient r and each with distribution N(0,1). Say 25% of the X values are missing completely at random, so that the correlation inferred on the 75% of cases that are complete provides an unbiased estimator of the true correlation. Replacing the missing values with values imputed solely from what is known of X just means using imputed values from N(0,1) that are uncorrelated with Y. Including them will be expected to reduce the correlation coefficient to 0.75r in the sample that includes imputed points, i.e. to give an estimate biased toward zero. If instead the missing X's are each replaced with a value imputed from the complete-case probability distribution conditional on *both* X and Y then the imputed x is of the form $ry+(1-r^2)^{1/2}\varepsilon$, where $\varepsilon$ is an N(0,1) random variable uncorrelated with X and Y. The correlation on the resulting sample is then expected to be r, an unbiased estimate of the true value.

In a more relevant case there are several predictor $X_i$'s with some correlations, not just one X. For simplicity we still assume that data are missing completely at random so that the correlations found in the complete cases are unbiased estimators of the true correlations, and assume multivariate normality so that the covariance matrix contains all the information. Imputing a missing $x_1$ via linear regression using the probability distribution for $X_1$ conditioned on both Y and the other $X_i$'s will still not bias the coefficient estimates since the imputed distribution shares all covariances with the complete-case distribution, by construction of linear regression. Imputing the missing $x_1$ via a probability distribution conditioned only on the other $X_i$'s would reduce the magnitude of the projection of Y onto the X predictor space, i.e. reduce $R^2$ of the model fit, just as in the toy one-dimensional case. Thus omitting outcome variables from the conditioning biases estimates to become less predictive even in the simplest, least problematic cases.

Choosing Variables to Include in Causal Models

A recent methods paper(Walsh et al. 2021) attempts to introduce PRPER readers to some of the issues involved in inferring causation from correlations, focusing on bias that can result from omitting variables. Although much of the verbal introduction is correct, the rules about which variables should be controlled for are stated incorrectly and the examples used undercut the message that one needs to carefully define causal questions before determining the correct combination of correlations needed to infer a causal effect with minimal bias. The title already conveys that bias arises from *omitting* variables from a model, but as we have seen bias can equally well arise from *including* inappropriate variables in a model.(Pearl, Glymour, and Jewell 2016; Hernán and Robins 2020; Greenland 2003) (Here and below "including" means controlling for by inclusion in a regression model, the usage in ref. (Walsh et al. 2021).)

The following general rule is stated(Walsh et al. 2021):

> "A variable included in a model will be biased by an omitted variable if the following two conditions are met:
> (1) the correlation between the omitted variable and the included variable is nonzero, and
> (2) the "true" effect of the omitted variable on the dependent variable is nonzero. "

This rule is simply untrue, since it would often imply that a causal model should include mediators and colliders. (Rohrer 2018) *Following this rule would allow one to show that essentially any causal effect is zero by including enough mediators in the model.*

The introduction (Walsh et al. 2021) promises to give examples using "explanatory" models as distinct from "predictive" models for "testing causal hypotheses". Nevertheless, the examples (Walsh et al. 2021) consistently contrast different "predictive" models without specifying causal patterns. Coefficients in some models are described as "biased" as compared to coefficients in others. This reflects a misunderstanding of what "bias" means. Predictions can be improved by adding parameters, but the resulting predictive coefficients for some suspected cause may either become more biased or less biased estimates of the causal effects of *intervening* on that cause, depending on the relations among the different variables in a causal diagram. (Pearl, Glymour, and Jewell 2016; Hernán and Robins 2020; Rohrer 2018)

The paper goes on to discuss various cases in which adding a variable to a multiple regression model makes either large or small changes in the coefficients of other variables. (Walsh et al. 2021) Throughout, the assumption is made that the extra variables should always be used when they make large changes, in order to approximate the "true" model (quotes in original). (Walsh et al. 2021) The question investigated is "whether omitting a particular variable will lead to

bias…" (Walsh et al. 2021) but the other half of the question, whether *including* a variable will lead to bias, is simply not mentioned. As we have seen, this approach is fundamentally mistaken because inclusion of colliders or mediators as covariates often systematically biases the relevant model coefficient away from the desired causal coefficient.(Pearl, Glymour, and Jewell 2016; Hernán and Robins 2020; Glymour, Zhang, and Spirtes 2019; Glass et al. 2013; Varian 2016; Foster 2010; Gangl 2010; Keele 2015; Greenland 2003; Rohrer 2018).

The first example given (Walsh et al. 2021) follows the same logical pattern as the smoking→tar→ cancer example. It concerns improved outcomes for males, compared to females, when labs are switched from real-world to virtual reality (VR), i.e. the difference between VR_benefit for males and VR_benefit for females. (Walsh et al. 2021) The wording of ref. (Walsh et al. 2021) gives the impression that the causal question is the effect of gender on VR_benefit. It is not entirely clear to what interventions knowing the causal effect of gender on VR_benefit would be relevant since it is unlikely that anyone deciding on some gender-determining action would be much concerned with the effects of their decision on optimal physics lab modes. As a purely predictive relation it could be relevant to deciding which students might benefit from which lab mode, but for purely predictive relations one needs only conditional probabilities, not a causal model.

A raw comparison of the distributions of VR_benefit for males and females suffices for predicting the effects on these groups. The paper claims that the model should include another variable, video game experience, which correlates very strongly with gender and strongly with VR_benefit. (Walsh et al. 2021) The model including gaming gives results that are contrasted with any "conclusions about biological or sociological differences in men's and women's ability to learn from VR", (Walsh et al. 2021) although it is unclear why gaming is not considered a "sociological difference".

The diagram in Fig. 1 shows a common-sensical causal pattern, in which being male leads, in the particular social context, to video gaming. Being male can lead to VR_benefit by the video game path or by other paths so the causal effect of being male on VR_benefit is just given by the unconditional regression coefficient in this simple linear model. Inclusion of gaming as an independent predictor biases the estimate of the causal effect of maleness on VR_benefit because it eliminates one of the causal paths for that effect. Ref. (Walsh et al. 2021) suggests "running the analysis with video game experience in lieu of gender", which corresponds to erasing not only the arrow from gender to gaming but also the arrow from gender to VR_benefit, in effect treating gender as an irrelevant marker for the gaming.

Which coefficients in the common-sense model are relevant depends on what action is being considered. Is the potential action choosing whether to offer VR labs at all, choosing which students to encourage to take them, encouraging students to play video games, or something else? To give a not very serious example, if one wants to know whether students should be encouraged to play video games to prepare for VR labs, then one would want to know to what extent gaming is an actual mediator on the gender→VR_benefit path or just a marker for gender. Because the correlation between maleness and video gaming is said to be very high(Walsh et al. 2021), that may be hard to disentangle. Nevertheless, clarity about what actions are under consideration would help to guide which coefficients in which diagrams are relevant.

Although the paper is not proposing anything that would actually affect choices about interventions on gender, the method used for estimating the effects of gender would consistently give incorrect causal coefficients for treatments on which interventions are possible. The causal question addressed is precisely analogous to the more serious one of finding whether inflammation is an actual mediator of SARS-CoV-2→ death, which determines whether intervening with dexamethasone is useful. For an educational policy example, inclusion of time spent viewing lectures in a multiple regression model would bias the estimand of the effects of two different video lecture styles on exam scores, because that variable would be a mediator on the path from style to scores. Viewing time would become relevant only if there were another way of intervening on it.

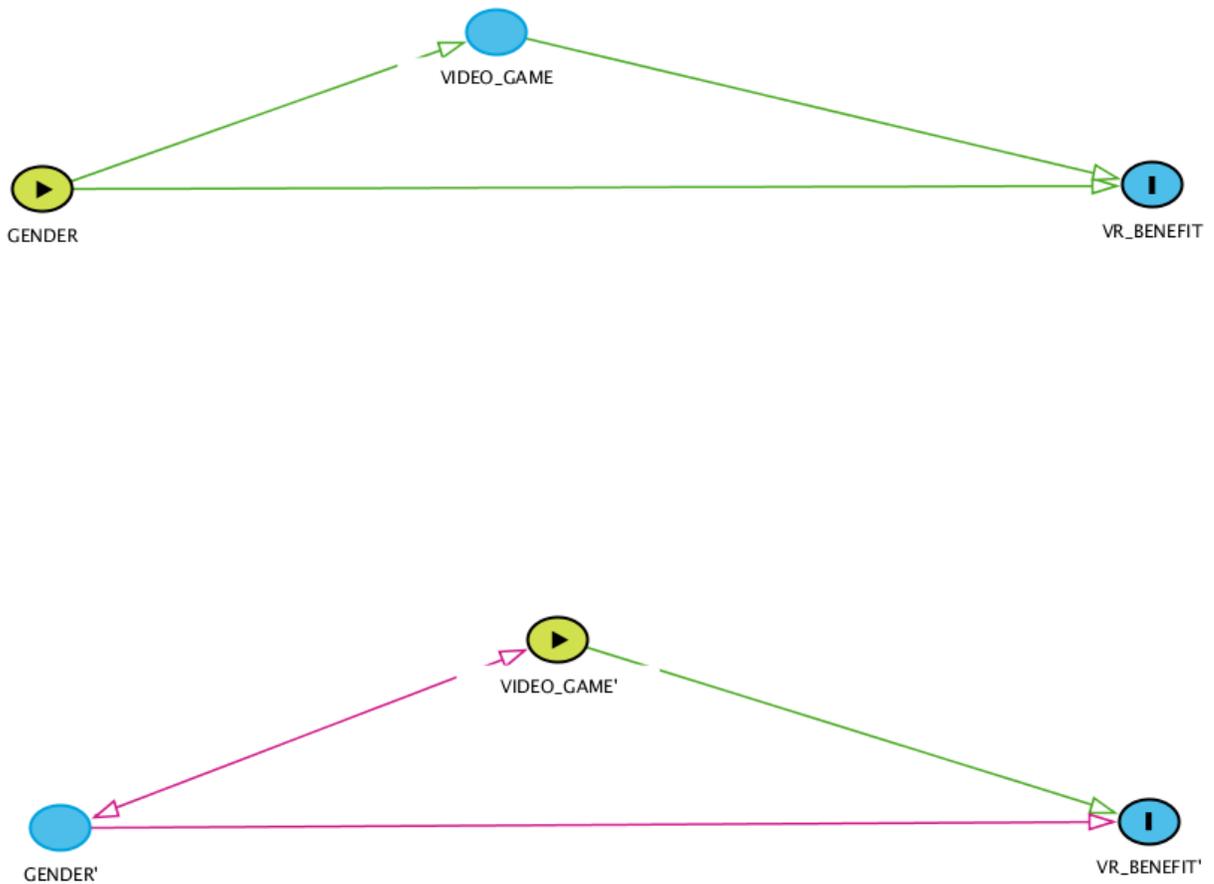

Fig. 1. The top figure represents a common-sense view of a causal relation between gender, gaming, and response to virtual-reality labs. Gender is an exogenous variable, i.e. its causes are not shared with other variables because at the time of the study they were almost always a quasi-random conception lottery. The Markov-equivalent second graph is the one implicitly used to obtain the Gender→ VR_Benefit coefficient in which gaming and gender are treated as correlated variables neither of which causes the other. One interpretation given by (Walsh et al. 2021) would also drop the arrow from gender to VR_benefit, giving an inequivalent graph that would in principle be distinguishable via observed correlations.

Identifying mediators and specifying potential outcomes

A recent PRPER paper examined the causes of admissions to selective graduate programs in physics, focusing on the roles of Graduate Record Exams in physics (GRE-P) and undergraduate grade point averages (UGPA), employing a structural equation model (SEM) analysis (Bollen and Pearl 2013) of the causal role of these variables and others on admissions. (Young and Caballero 2021) The causal theme is to see what effects dropping GRE-P would have on admissions, and in particular whether GRE-P is helping some students be accepted who might otherwise be overlooked. I shall discuss here only problems in the translation of statistical claims to causal claims, not those in the conversion of data to statistical claims. These problems include an incorrect rule for identifying causal mediators and unrealistic framing of the potential outcomes for the interventions under consideration.

The SEM analysis emphasized testing for mediation. (Young and Caballero 2021) Although the verbal definition of causal mediation is presented correctly, the mathematical method presented for identifying mediators treats one causal variable (e.g. GRE-P) as mediating a path from another cause (e.g. UGPA→admission) whenever the "mediator" can be expressed in terms of the other cause plus a random component, as explicitly shown in their Equation 2. (Young and Caballero 2021) The logistic regression SEM analysis assumed that GRE-P mediates UGPA→admission and concluded that the mediation effect was large because the dependence on UGPA of the logit for the admissions probability changed substantially when GRE-P is held constant. (Young and Caballero 2021)

Nevertheless, it is evident that the choice of which correlated variable, if either, to call a mediator cannot be based just on the correlation itself, since that relation is essentially symmetrical: the least-squares estimate of each normalized variable is simply the correlation coefficient r times the other normalized variable, plus a variable that's orthogonal to the predictor variable and has variance $(1-r^2)$. This familiar symmetry is just the simplest case in which different causal graphs have exactly equivalent predictions for correlations. (Pearl, Glymour, and Jewell 2016; Richardson 2003) Following the procedure of (Young and Caballero 2021) would produce evidence of "mediation" in all cases where two causal variables are correlated, regardless of whether any actual causal mediation exists. That procedure confuses

1/31/22 12

the estimation of coefficients for a known causal diagram with the discovery of the proper causal diagram.

To give a graphical illustration of the issue, I turn to the SEM on the effects of "Race", Fig. 13 of (Young and Caballero 2021), reproduced here as Fig. 2. (In ref. (Young and Caballero 2021) "Race" stands for an aggregate approximately meaning the same as "under-represented minority".) The correlations were interpreted to show that GRE-P mediates the causal effects of GPA on admission, and thus also the effects of "Race" on admission via UGPA. What the correlations themselves show, however, is that "Race" has very little correlation with GRE-P (not statistically significant in the sample) when UGPA is held constant. No discussion is included as to why the relation UGPA→GRE-P is assumed. (Young and Caballero 2021)

The mistaken mathematical conception of mediation leads to at least one strange implication. A literal reading of the original SEM, taking its causal diagram seriously, would say that the way to get rid of the negative effect of Race on Admit would be *to eliminate GPA*- not just to drop it as an admissions factor, which would still allow a negative Race effect to flow through it to GRE, but to actually eliminate it, perhaps by abolishing grades. That would almost entirely block the path for the negative causal impact of Race on Admit, since almost none of that impact bypasses GPA in that SEM.

A more conventional view might be that UGPA and GRE-P are correlated because both are affected by a variety of shared factors, e.g. diligence in studying physics. This alternative is illustrated by the second graph in Fig. 2, which is Markov equivalent to the first graph and therefore exactly equally consistent with the correlations. (Pearl, Glymour, and Jewell 2016; Richardson 2003) This modified SEM does not have any peculiar (presumably unintended) implications about what effect eliminating UGPA would have on GRE-P. Although for the most part UGPA is obtained before GRE-P, that has no implications for the causal order, if any, between the latent variables underlying them.

The correlations encoded in the original SEMs have other, less peculiar, policy implications than the SEMs themselves. The original SEM and the revised one are Markov equivalent, so they share the same correlations between Race and the variables UGPA and GRE-P, i.e. -0.48 and-0.31, respectively. (-0.31 is simple the sum of the effects on the direct and mediated paths from Race to GRE-P in the original SEM, -0.094-0.480*0.445.) That means that replacing GRE-P with more weight on UGPA could plausibly have *decreased* the admission of under-represented minorities in this cohort, a conclusion that matters for the policy choices under consideration. This simple conclusion is not included in the paper. For gender, in contrast, such a substitution would favor females, as can be seen from the coefficients in Fig. 12 of ref. (Young and Caballero 2021).

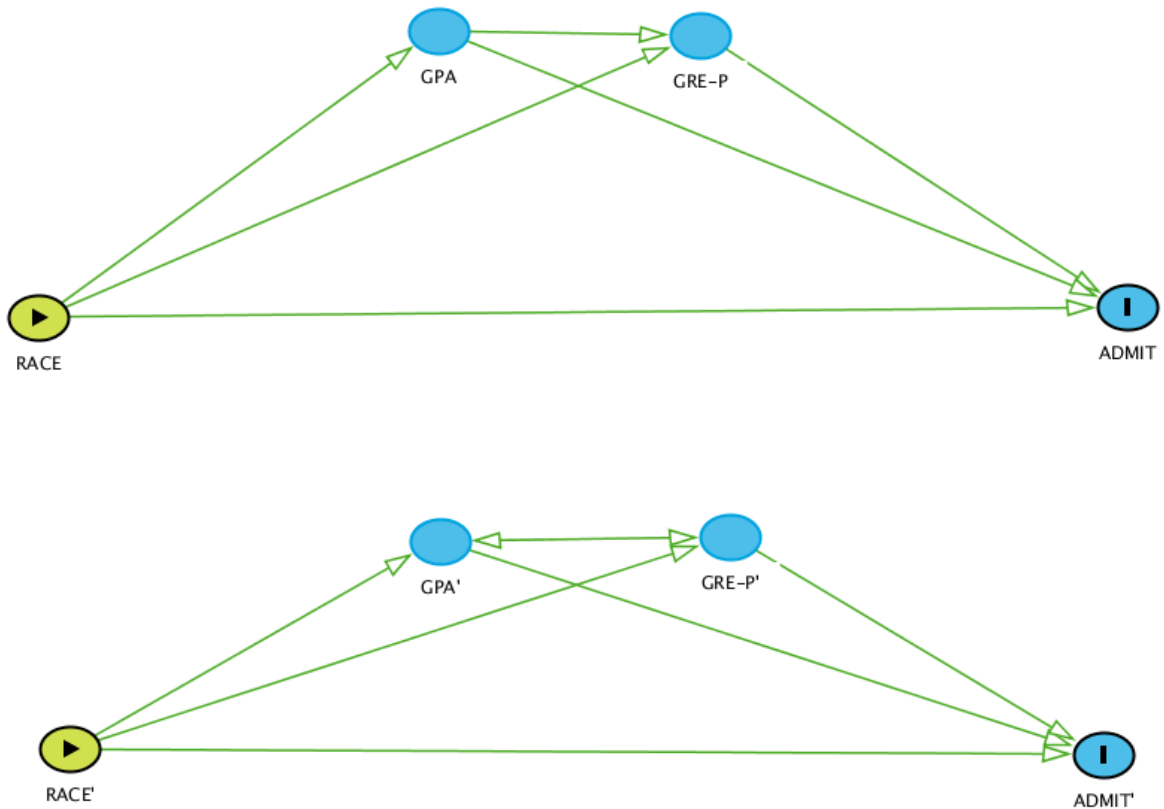

Fig. 2. The SEM graph used to describe the causal effect of "race" on admissions, as given in (Young and Caballero 2021) on top, and using a Markov equivalent and seemingly more plausible graph on the bottom. Bidirectional arrows stand for causation by exogenous unmeasured variables, not two-way causation. The latter graph avoids the false implication that dropping GRE-P would reduce the effects of "racial" differences in GPA on admissions.

Another fundamental issue arises in the framing of the causal effects, i.e. potential outcomes, of changing admissions criteria. Unsurprisingly, ref. (Young and Caballero 2021) shows that admission probability tends to increase with higher UGPA and GRE-P. The paper emphasizes that "more applicants could be penalized for having a low physics GRE score despite a high GPA than could benefit from having a high physics GRE score despite a low GPA." (Young and Caballero 2021) The net effect on acceptance of changes in acceptable levels for different criteria, however, depends on the ratio of the hypothetical incremental changes for those two criteria and thus cannot be determined without specifying that ratio. Ref. (Young and Caballero 2021) does that specification via arbitrary definitions. More fundamentally, framing the outcome as the total number of accepted applicants does not consider the actual potential outcomes, in which the total number of accepted applicants who can attend graduate school is not set by admission criteria but by limits on funding, mentorship, and job openings. Any change of admission criteria may be judged on a variety of grounds, but these do not include effects on the total number of new graduate students, which can only be changed by other methods. Changing criteria changes which students are accepted, not how many. The comparison of net acceptance rates for different criteria may, however, make sense if the implicit goal is to increase the number of domestic students by reducing the number of international students, who were omitted from the analysis. (Young and Caballero 2021)

Implications for use of GRE scores: collinearity, collider bias, and imputation bias

Many physics and astronomy departments have dropped or are considering the possibility of dropping use of GREs for admission, dropping either just GRE-P or both GRE-P and the more widely taken quantitative GRE, GRE-Q.(Chawla 2020; Young 2020) Estimating the causal effects of these actions is therefore worthwhile. Some of the pathways by which these choices might affect various outcomes (e.g. by effects on both graduate and undergraduate curricula and grading) are essentially unmeasurable without trying the experiment. Other pathways, such as the potential performance of students who are not currently admitted, may be roughly estimated via model-based extrapolation from results on how well GREs predict graduation in the currently admitted cohort. (Miller et al. 2019; Weissman 2020b; Miller et al. 2020; Weissman 2020a; Verostek, Miller, and Zwickl 2021) Here I address causal issues in a recent PRPER paper(Verostek, Miller, and Zwickl 2021) that uses this latter approach. Although ref. (Verostek, Miller, and Zwickl 2021) uses somewhat vague language to describe its aims (e.g. asking which metric "offers the most insight"), in context(Miller and Stassun 2014; Chawla 2020; Young 2020) there is no doubt that the treatments under consideration are to drop use of GREs in graduate physics admissions. Unlike for the previous two papers(Walsh et al. 2021; Young and Caballero 2021), each of which was characterized by a small number of core errors in causal reasoning, for this paper I will discuss more conceptually disparate errors, united by sharing a sign of their effect in biasing causal estimates.

A causal diagram approximately representing the actual admissions policy choices to be made is shown in Fig. 3. Different criteria may be used or not used in choosing whom to admit. These criteria include the ones that are relatively easy to tabulate and study (UGPA, GRE-Q, and GRE-P) and various other factors that are used by most departments(Potvin, Chari, and Hodapp 2017) but are harder to quantify and tabulate across schools and thus comprise out-of-model-predictors (OOMP). The effect of choosing to use or not use a criterion is conditional on choices for the other criteria, since inclusion of a highly redundant predictor has little effect, and redundancy depends on what else is used.

To determine the effects of dropping both relevant GREs (Q and P) on the predictability of graduation, one should compare predictions using those two GREs with predictions excluding them but otherwise the same. That direct comparison for graduation rates is not made in the PRPER paper(Verostek, Miller, and Zwickl 2021) or its predecessors(Miller et al. 2020, 2019), but we shall see that it may be estimated from the data presented.

Fig. 4 shows a causal diagram roughly representing the key points of the model for what causes an individual student to be more or less likely to graduate. Coefficients from this model can then be used implicitly or explicitly to estimate the effects of using different metrics in the policy diagram of fig. 3. Fig. 4 encodes some assumptions, probably not quite exactly true but not especially controversial. One is that whether a student is admitted and if so to what rank of department depends on the explicit metrics available to the admissions committees, (Young and Caballero 2021) not on the traits that they partially measure. Another is that the metrics only partially reflect some unspecified traits of the applicants that make them more or less likely to graduate. Correlations between metrics arise because they measure overlapping traits. I arbitrarily use four such traits in fig. 4, enough to allow linear independence of their different combinations in the four metrics.

A causal diagram is included as Fig. 5 in ref. (Verostek, Miller, and Zwickl 2021), but since it includes only a single metric, it does not capture that each causal effect in our Fig. 3 is conditional on which other metrics are also used. That dependence approximately corresponds to the dependence of the predictive coefficients on which other predictors are included in a model for individual students like Fig. 4. Thus the diagram used in (Verostek, Miller, and Zwickl 2021) obscures the key point- that dropping either GRE substantially raises the coefficient and the statistical significance of the other GRE for graduation probability.(Weissman 2020b, 2020a) The effects of dropping one predictor on the other's coefficient are later given (Verostek, Miller, and Zwickl 2021) but only for a different outcome, graduate GPA (GGPA), for which the effect is small because GRE-Q adds little to its prediction, unlike for graduation.

To estimate the causal coefficients of the policy-choice model one needs first to correctly estimate the coefficients and statistical uncertainties of the predictive coefficients of the individual-level model. Ref. (Verostek, Miller, and Zwickl 2021) gives coefficients and confidence intervals for the coefficients of GRE-P and GRE-Q for predicting graduation within the stratum of enrolled students within a model including UGPA and various demographic predictors. Within the main text, it is stated at least 12 times that neither of the separate coefficients meets the conventional cutoff for statistical significance. (Verostek, Miller, and Zwickl 2021)  In the Supplement, however, we find that when a somewhat less biased method is used for imputing missing data, both coefficients actually do pass the standard "significance" criterion. (Verostek, Miller, and Zwickl 2021) (Here I do not mean to endorse the common null-hypothesis-significance testing approach of making qualitatively dichotomous interpretations of small differences in coefficient p-values (Amrhein, Greenland, and McShane 2019), but merely go along with it to focus on other issues.)

In ref. (Verostek, Miller, and Zwickl 2021) as well as in its predecessors(Miller et al. 2020, 2019) a substantial amount of missing predictive data was filled in using a previously unspecified(Miller et al. 2019) multiple imputation method. Ref. (Verostek, Miller, and Zwickl 2021) now points out that in its main text as well as in the preceding work the imputation model used omitted outcome variables, claiming that "the imputation approach presented here is theoretically sound".  The Supplement (Verostek, Miller, and Zwickl 2021) expresses surprise that the "counterintuitive" inclusion of outcome variables results in less biased estimates, based on concern that "Employing a model of data imputation that uses the outcome variable to predict missing values of the independent variable may seem like a self-fulfilling prophecy, guaranteeing a relationship to exist between them."  Nevertheless, the Supplement (Verostek, Miller, and Zwickl 2021) concedes that "research suggests that including all variables, including the outcome variable, in the imputation model in fact tends to produce less biased results."  We have seen, however, that one needs only algebraic inspection of the simplest models to reveal that imputation methods that omit outcome variables from the conditioning are not "theoretically sound" because they introduce systematic bias. (This issue may provide a clue as to how proficiency in beginning algebra, as measured by GRE-Q, could predict research performance.)

The Supplement of ref. (Verostek, Miller, and Zwickl 2021)  now gives imputations including conditioning on GGPA. (Verostek, Miller, and Zwickl 2021) Even that partial correction suffices to make GRE-P and GRE-Q "significant" predictors of graduation in the U.S. cohort. It appears, however, that graduation itself was still not included in the imputation method of the Supplement, so the results would still be biased toward weakening the model's predictive power, especially for that outcome. Estimating from blow-ups of figures in the main text and the Supplement as well as from the decrease in p-values and the partial data given on effect size for GRE-P, just conditioning on GGPA raised the sum of the standardized coefficients of GRE-P and GRE-Q for graduation by ~8%. The standardized coefficient of UGPA went up ~50%.

Insufficient data are given to tell how much the coefficients for graduation would go up if that imputation were also conditioned on the more relevant outcome, graduation itself. One may get a rough idea from the effect of GGPA conditioning on the coefficients for predicting GGPA. When GGPA was used in the imputation conditioning $R^2$ for the model increased from 0.11 to 0.17 and the coefficients of UGPA and GRE-P for GGPA went up ~35%. (Verostek, Miller, and Zwickl 2021) (The coefficient of GRE-Q for GGPA was very small and not significant in either imputation. (Verostek, Miller, and Zwickl 2021)) It would be reasonable to guess that the sum of the GRE coefficients for graduation would also go up another 10% to 30% if graduation were properly included in the imputations for that model. The actual value could be calculated easily by the original authors.

Some of the graduation outcome data were also missing because at the end of the data window some students had been enrolled for five years but had not yet graduated. (Verostek, Miller, and Zwickl 2021) Although it was stated that roughly 95% of these were likely to graduate, the model treated their graduation rate as 100%, in effect substituting five-year survival for graduation as the outcome. (Verostek, Miller, and Zwickl 2021) A less biased result would be obtained by using imputation based on the 95% estimate. The probable effect of this fairly small unnecessary bias, like that of the larger errors, is to understate the predictive power of GREs.(Weissman 2020b, 2020a)

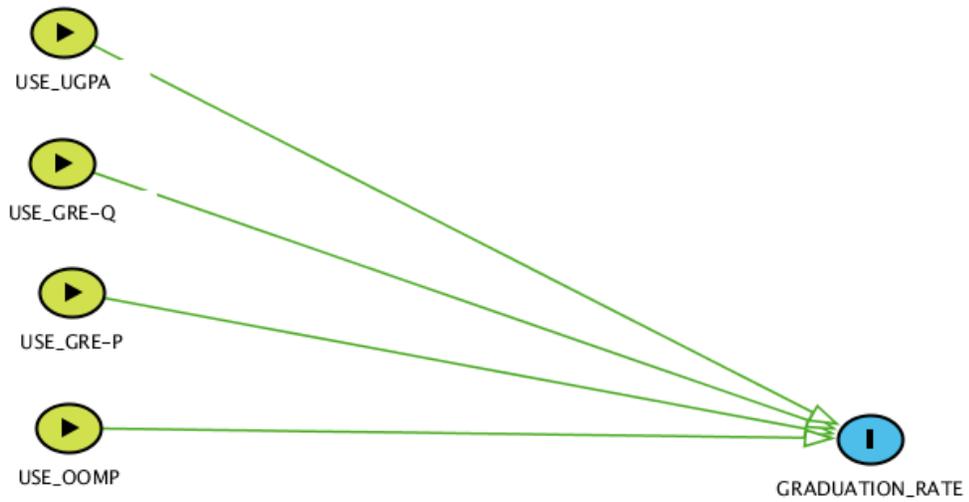

Fig. 3. This graph illustrates the effects of possible choices of admissions criteria on the net graduation rate for a particular program. Out-of-model predictors (OOMP) are available to admissions committees but not to subsequent modelers.

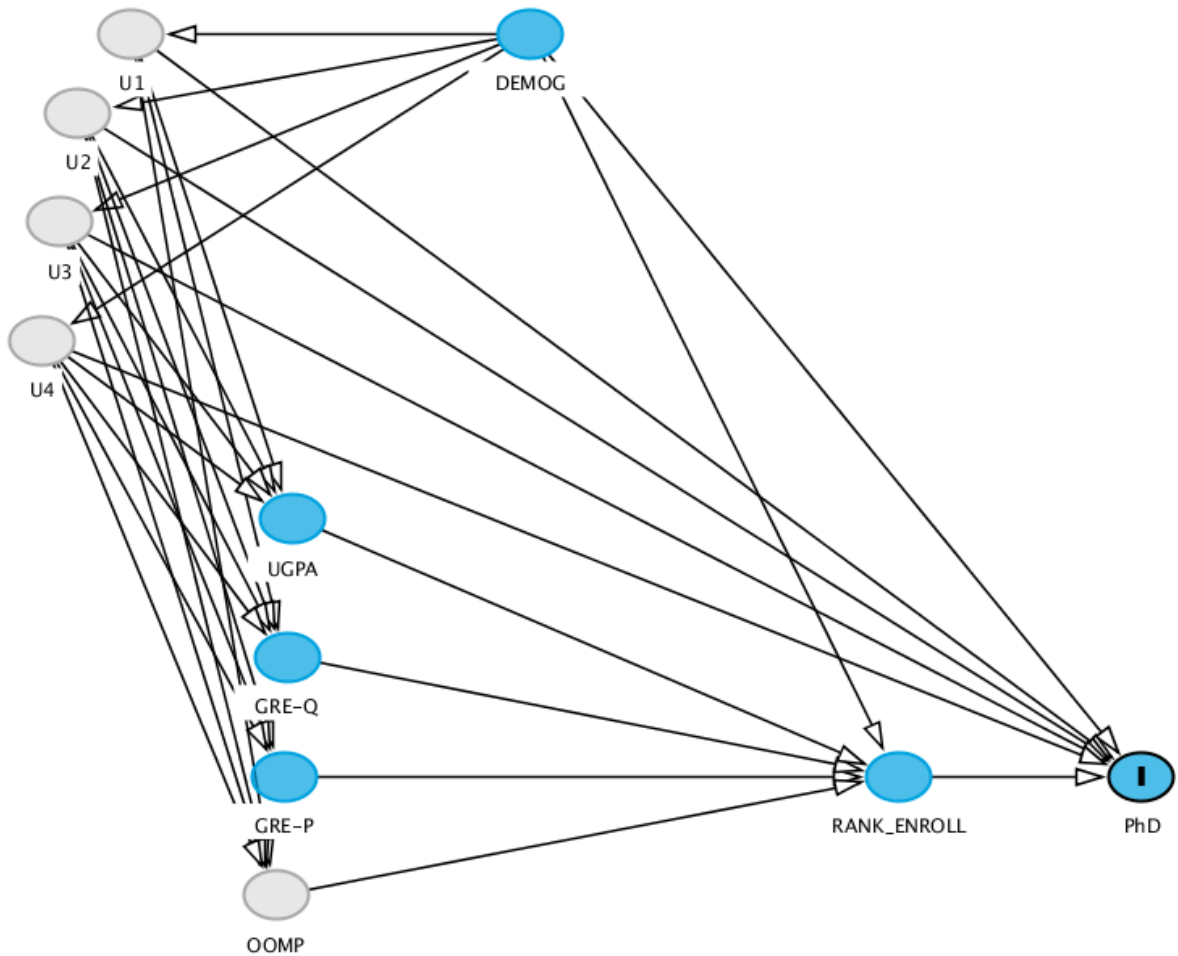

Fig. 4. This graph illustrates how the variables available to the admissions committees and the subset available for the overall model are related to the probability of graduation. It contains some explicit assumptions, e.g. that the causal effects of the variables with no direct arrows to PhD are negligible except as mediated by the necessary step of getting into at least one program, even though those variables can be important markers of traits that do affect graduation rates. The correlations among the predictor variables arise because they share underlying causes. Since there are no data on students who were not admitted to any programs, some bias from stratifying on the collider "rank_enroll" is unavoidable. Stratification into rank tiers or into the narrower strata of individual programs increases that bias.

Once the coefficients relevant to fig. 4 have been established, one must estimate their implications for the coefficients of the policy choices of fig. 3. Since GRE-Q and GRE-P are highly correlated, collinearity effects make the separate point estimates and confidence intervals for their coefficients not directly relevant to the question of whether dropping both predictors would significantly reduce predictive power. *Estimating the effects of dropping one or both predictors requires a calculation that uses both the coefficients and the correlations among the predictors.* A previous analysis(Weissman 2020b, 2020a) showed unambiguously that the predictive power for graduation would fall significantly if both GREs were dropped, an effect of more than four times the standard statistical error even using the improperly imputed estimates. We shall see that consideration of the information in the new PRPER paper (Verostek, Miller, and Zwickl 2021) will give the same conclusion but with larger effect sizes.

Even correct use of within-sample predictive coefficients would still give somewhat biased estimates of how much predictive power would be lost by dropping each predictor, because the data are necessarily restricted to those students who were admitted and then enrolled. (Weissman 2020b, 2020a) As we see in Fig. 4, graduate admission is a collider between the model variables (UGPA, GRE-P, GRE-Q, demographics) and OOMP, a summary of other admission criteria used by almost all programs. (Potvin, Chari, and Hodapp 2017) Some collider stratification bias is thus inevitable.

Table V of ref. (Verostek, Miller, and Zwickl 2021) allows an estimate of the magnitude of such collider effects since it gives the results of the most recent analysis and that of a previous one in which extra collider stratification was introduced by dividing the graduate programs into three tiers of rank.(Miller et al. 2019) (The comparison is very slightly complicated because the new estimates appear to use a probit link function and the old ones use a logit link function, but that should have very minor effect on this comparison. Also the new numbers use only ~80% of the old sample, but we are not given any indication that that change affects the comparison. (Verostek, Miller, and Zwickl 2021)) The new less-stratified graduation predictive  effect sizes for UGPA, GRE-P, and GRE-Q were increased from those found in the  more-stratified model by factors of 1.7, 1.2, and 1.6, respectively. This substantial change illustrates the importance of collider bias. These factors are far more precise than one might guess from a raw combination of the separate confidence intervals on the more and less stratified model results, because both models use the same set of partially random outcomes. The actual narrow confidence intervals for the factors could be determined by a pairwise bootstrap calculation using the actual data sets.

We may apply these correction factors to my previous calculation of the effect sizes for the logit of graduation probability as a function of UGPA and equal-weight GRE sum.(Weissman 2020b, 2020a) Based on the highly stratified results, the logit effects on going from the 10$^{th}$ to 90$^{th}$ percentile among the U.S. test-taking cohort, holding other predictors constant, were previously found to be 0.60 and 0.72 for UGPA and equal-weight GRE sum, respectively.(Weissman 2020b, 2020a) After correction for the excess stratification of that model each logit effect is 1.0, coincidentally. I suspect that the remaining stratification bias from the necessary absence of OOMP in the newer less-stratified results is not as big as the difference between them and the highly stratified earlier results, in part because the remaining stratification bias is likely to be partly balanced by some confounding bias.(Greenland 2003; Weissman 2020b; Rohrer 2018)

Using the new imputation results, partially repaired by including GGPA conditioning, would raise those estimates to 1.5 and 1.1, respectively, for UGPA and GRE_sum. Fully repairing the imputation method would likely raise both effects substantially more. It would be good to see the actual result of that straightforward calculation, preferably on the full data set used in ref. (Miller et al. 2019).

Estimating the causal effects of admitting students who are now excluded because of low GRE scores or because they did not take any GREs requires extrapolation of results from students who are currently admitted to that broader cohort. Extrapolation involves use of some model link function, e.g. either logit or probit, to convert predictors to probability. This extrapolation is uncertain, especially for GRE-Q, for which the range among all the students who take the test and are interested in going on in physics is more than twice as large as in the enrolled group. (Miller et al. 2020) Range restriction is much less severe for GRE-P and UGPA. (Miller et al. 2020) Data on the enrolled group can provide at least some hint as to whether that extrapolation is problematic, e.g. by seeing if outcomes for the bottom quintile of those enrolled fall below expectations from the overall model fit. Since extrapolation to the full U.S. cohort, even via the linear logistic model, predicts bigger effects for GRE-Q than for GRE-P (Miller et al. 2019), it would be particularly important to see a quintile breakdown for GRE-Q. The Supplement to ref.

(Verostek, Miller, and Zwickl 2021) provides such graduation results for quintiles of UGPA and GRE-P but omits the potentially more important results for GRE-Q.

We can now turn to the question of whether dropping just the GRE-P while retaining GRE-Q would significantly reduce predictive power. According to the analysis in the Supplement(Verostek, Miller, and Zwickl 2021), GRE-P has a significant predictive coefficient in the full model. From results(Miller et al. 2020) for the Akaike Information Criterion(Akaike 1974), dropping both GREs would give a much worse model overall and within the U.S. cohort but dropping just GRE-P only gives a marginally worse U.S. model. The loss of predictive power should be somewhat less than the point estimate of the GRE-P predictive coefficient might lead one to expect. (Weissman 2020b, 2020a) The reason again is that once one predictor is dropped, the coefficients of the other predictors change. GRE-Q and UGPA, both correlated with GRE-P, would take up some of the slack left by dropping GRE-P.(Weissman 2020b) That's also why if one GRE is dropped, the incremental predictive power obtained from the remaining one is increased, so that dropping the second GRE gives a *larger* loss than its coefficient in the full model would suggest. (Weissman 2020b, 2020a)

The primary motivation for examining the effects of using GREs on admissions is stated to be concern for under-represented minorities (URM, approximately corresponding to "Race" in ref. (Young and Caballero 2021)) and for females. (Verostek, Miller, and Zwickl 2021) The effect of dropping GREs on URM admissions will depend on the balance of which other factors are used to replace them. Although ref. (Verostek, Miller, and Zwickl 2021) does not specify what combination would be used, it emphasizes the superiority of UGPA as a predictor. We have seen, however, that in the cohort studied in ref. (Young and Caballero 2021) the standardized prediction coefficient for "Race"→UGPA is *more* negative than that for "Race"→GRE-P. (The coefficient for GRE-Q was not given.) Thus not only would dropping GREs have a greater negative effect on graduation probability than suggested in ref. (Verostek, Miller, and Zwickl 2021), but it could also have a negative effect on URM admission if more weight is placed on UGPA.

On the other hand, the GREs (especially GRE-P) , unlike UGPA, do systematically underpredict both female graduation rate and GGPA. (Verostek, Miller, and Zwickl 2021)  Female admissions could be increased by dropping or de-emphasizing them. The same goal could also be achieved by gender-norming those scores, without losing the significant information provided by the GREs. It seems likely that something like that norming was already being done in the period for which data were collected, since the overall female GGPA is insignificantly less than the male GGPA. (Verostek, Miller, and Zwickl 2021) Without something like gender normed admissions, the female GGPA would have been expected to be higher than the male GGPA, since it is significantly higher than predicted by a UGPA-GRE model. (Verostek, Miller, and Zwickl 2021) Fig. 12 of ref. (Young and Caballero 2021) confirms that the standardized coefficient for predicting the admissions probability logit for females in a model controlling for GRE-P and UGPA was in fact large and positive.

Much of ref. (Verostek, Miller, and Zwickl 2021) is devoted to a "mediation analysis" in which GGPA is treated as a potential mediator on the paths from UGPA, GRE-Q and GRE-P to graduation. The predictive effects of UGPA and GRE-P for graduation are described as almost entirely "mediated" by GGPA because controlling for GGPA removes almost all of the dependence of graduation probability on UGPA and GRE-P in the model. Just as in the mediation cases we examined above, however, that result does not show whether GGPA actually mediates the causal effects on graduation of the traits measured by UGPA and GRE-P or merely serves as another marker for those traits. Only to the extent that GGPA actually mediates the causation or serves as a marker for true mediators, as opposed to being a marker for pre-existing traits, would interventions that raise it (e.g. better classroom teaching) increase graduation rates. (Interestingly, controlling for GGPA scarcely changes the dependence of graduation probability on GRE-Q in the model, suggesting that completing a PhD partly depends on some traits measured by GRE-Q that would not be changed much by changing classroom teaching.)

UGPA and GRE-P are described as providing only "indirect prediction", an unusual cross between causal and predictive terminology. The use of causal terms, "mediation" and "indirect", in describing a purely predictive relation, is somewhat misleading. The pre-admission predictors are no more or less predictive of graduation than they would be if post-admission GGPA had not been subsequently recorded.

1/31/22                                                                                                                                                        25

The methodological errors in ref. (Verostek, Miller, and Zwickl 2021) are important in themselves, but in this case, the substantive conclusions are also important, so a summary of the key corrected conclusions may be useful. Unlike UGPA, GREs remain useful in the full cohort, not just the domestic subset. Within the group of all applicants, GREs provide more incremental predictive power for PhD attainment than does UGPA. Within the cohort of U.S. applicants' GREs provide about the same incremental predictive power as UGPA. *Within the cohort of domestic students who are interested in physics graduate school and who have the same UGPA, dropping both GREs would lose an odds ratio of at least a factor of three in estimating how the probability for graduation varies between high-scorers and low-scorers.* Dropping only GRE-P would lose substantially less predictive power than dropping both GREs.

Discussion

We have seen several major errors concerning causal inference in recent PRPER papers, including in explicit methods sections. The errors are directly consequential for estimating the effects of policy choices, including a major policy choice under active consideration. These errors reinforce the impression given by different errors in other papers (Weissman 2021) that the PER field needs better knowledge of causal inference techniques.

In all the papers discussed here and in my previous PRPER article(Weissman 2021), there was no clear specification of which actual interventions and consequences would be considered. Instead, implicit rather than explicit assumptions were used to justify applying more or less causal language to collections of correlations. Policy recommendations were then intuited or implied, leaning on causal interpretations of some conveniently chosen correlations. Although a similar description might apply to much other social science research, especially from before about 1980, many fields have now moved well beyond that.(Pearl, Glymour, and Jewell 2016; Hernán and Robins 2020; Glymour, Zhang, and Spirtes 2019; Glass et al. 2013; Varian 2016; Foster 2010; Gangl 2010; Keele 2015; Murnane and Willett 2011; Rohrer 2018; Imbens and Rubin 2015)  PER should join them.

Perhaps the most important lesson is that one should start with an explicit question about reasonably well-defined potential consequences of reasonably well-defined choices of potential

actions.(Robins and Weissman 2017) Even when a randomized controlled trial is infeasible, just trying to imagine one can help to clarify what question is being asked.(Hernán 2021) Although explicit assumptions are still needed to pare down the possibilities to something tractable, that starting clarity can go a long way toward guiding the type of causal graphs that are needed. Graphs then help to see what combinations of correlations provide the best estimate of the effects of possible actions.

Some procedural changes might help reduce the current problems. One might be to preregister protocols for studies.(Chambers 2019) That allows some refereeing of methods before data are gathered and before particular methods are closely tied to particular conclusions. That helps limit cherry-picking methods and data, reducing p-hacking and reverse p-hacking(Chuard et al. 2019). Even with pre-registration, however, one still needs a community of reviewers and readers who can distinguish between valid and invalid methods. Establishing that may require help from outside PER.

## **Acknowledgements**

I thank Jamie Robins for crucial help in understanding causal inference. I thank the PRPER editors for suggesting this project.